# A superfluid He$^4$ version of a test on QG vs CG: feasibility with demonstrated methods

Massimo Cerdonio* and Giovanni Carugno$^§$
INFN Section and University, Padova, Italy   June 4$^{th}$ 2021

* cerdonio@pd.infn.it   $^§$ carugno@pd.infn.it

Abstract
A field, which mediates entanglement between two quantum systems, must be of quantum nature. Attempts to witness this way quantumlike features of the gravitational field with tabletop experiments are actively studied recently, in particular by considering to look at two masses in a superposition in two locations, each in one interferometer. Entanglement intervention is probed when the interferometers are put side by side. If the masses interact only via Newtonian attraction, and still some degree of entanglement is found, than the gravitational field must be quantum like, or at least non-classical. The masses considered are *mesoscopic,* 10$^{-14}$ Kg to 10$^{-12}$ Kg, and in one proposal  Mach-Zehnder interferometry is considered. Liquid He$^4$ is superfluid below 2.17 K, and shows macroscopic quantum behaviour, in particular matter interferometry, as in the Superfluid He Quantum Inteference Device - SHeQUID. With its Josephson junctions as slits, the SHeQUID parallels a Mach-Zehnder. In this case the matter quantities involved are *macroscopic*, 10$^{-8}$ Kg. We propose and analize the feasibility of a scheme on the lines of the above, where the matter field is given by superfluid He$^4$, and the Mach-Zehnder's are two SHeQUID, put side by side. We find that the proposed experiment is feasible, using only well demonstrated methods and technologies, with no need to extensions beyond the current frontiers.

1.Introduction
The questions if gravity is classical down to fundamentals, CG, or if it emerges from a quantum theory of gravity, QG, or if it is just non-classical, but non necessarily quantum, has attracted much interest recently, with a wealth of proposals, analysis and debates, concerning table top tests [1-11].
      One must choose systems that show non-negligible Newtonian attraction, while no other kind of interaction is present, and that at the

same time can go into superposition states, specifically N00N states [8]. If some degree of entanglement occurs within parts interacting gravitationally, then this witnesses quantumlike features of the gravitational field, QG. The analysis in [3] reinforced the notion that table-top experiments, where matter fields are entangled by Newtonian gravitational interaction, are able to probe quantum features of gravity.

One proposal [1] is of particular interest here, because ours in a way parallels that. The proposal makes use of two equal masses in a state of superposition in two locations, as they propagate each in one of two Mach-Zehnder interferometers arranged side by side, with pairs of arms parallel. Each mass m enters separately the corresponding interferometer and interference effects at the outputs are observed. The masses interact only through their gravitational attraction. Should gravity have an underlying quantum nature, QG, the side-by-side beams would get gravitationally entangled, and the interference would be affected.

Consider the two Mach-Zehnder interferometers arranged as in Fig.1 of ref[1], and let us follow the notations therein, that is d1 is the separation between side-by-side arms, and d2 is one of the other arm separations. Before entering their respective final beam splitter, the masses acquire a path dependent phase increment, i=1,2

(1)     $\phi_i = m^2 \, (G/\hbar) \, (\Delta t / d_i)$

where m is the mass entering the interferometer, G the gravitational constant, h-bar the Planck constant, and $\Delta t$ the time spent by each mass in that arm.

In [1] a discussion is given on how the interference patterns are influenced by the $\phi_i$'s, if entanglement occurs or not. There are two extreme regimes. If the two masses are not entangled, they undergo ordinary interference. If, by contrast, maximal entanglement occurs, the interference is completely destroyed. Given the mass m, either there is no interferometer configurations - separations d1 and d2, time spent $\Delta t$ - in which the ordinary interference is destroyed, or in some configurations the interference is completely destroyed. If this can be reconducted to the relations of eq (1), then this constitutes a witness that the gravitational field mediating the interaction between the masses must be quantum at the fundamental level.

In [1] and other proposals of table top experiments [2,3,7,8,11] the masses involved would be *mesoscopic* solids $10^{-14}$ Kg to $10^{-12}$ Kg, or BEC's of up to about $10^9$ atoms. Quite recently it has been made a remarkable step forward in realizing in the lab [12] the scheme proposed in ref [2].

2. Concept of the superfluid $He^4$ experiment.
Here we propose a new concept, where the matter quantities involved are *macroscopic*. We take advantage of the fact that liquid $He^4$, a superfluid below $T_\lambda$ = 2.17 K, shows macroscopic quantum behaviour, in particular matter interferometry. In this case the interferometer is a loop of channels incorporating Josephson junctions, a macroscopic quantum interference device - SHeQUID (see below for details) - where the quantities of matter involved would be *macroscopic*, order of $10^{-8}$ Kg.

We propose a scheme which parallels that of ref [1], using SHeQUIDs as the Mach-Zehnder's. The system is operated at less than one mK below $T_\lambda$, where ideal Josephson effects in superfluid $He^4$ occur. The basics are robust, because such macroscopic quantum effects have been extensively observed and found in complete agreement with fundamental theories of macroscopic quantum phenomena in superfluids [13,14]. The essence of the proposed experiment is to look for destroying or mantaining the pattern of interference in the system of two SHeQUIDs positioned side-by-side.

3. The experiment with superfluid $He^4$.
The SHeQUID is a superfluid $He^4$ analog of the superconducting dc-SQUID. It is a well-studied case [14] - noticing ref [49] and ref [100] therein. In brief, two constrictions - the junctions - showing Josephson effects, are inserted in channels making a closed loop, where the superfluid flows. The system behaves as a matter interferometer. In the superconducting SQUID the sensing loop is sensitive to the flux of the magnetic field threading it. In the superfluid $He^4$ case the role of the magnetic field is taken by the vector of rotation in respect to the local inertial frame [15].

We propose to position side by side two identical SHeQUIDs, Fig.1. Each sensing loop has the geometry of a square of side length L and the channels have cross section $\sigma$. The channels are traversed by the superfluid component of density $\rho_s$ at temperatures less than one mK below $T_\lambda$. The two apparatuses have the planes of their loops residing in a vertical plane on Earth. Their channels of length L, lying respectively side-by-side, are parallel and horizontal at a distance d ≪ L.

The SHeQUID is considered the matter interferometer analogue of a Mach-Zehnder. This feature of the SHeQUID has been remarked in [16], where it is specified that the Josephson junctions constitute the beam splitters and the channels, in which the superfluid flows, constitute the arms. The superfluid in the two apparatuses must come from well-separated $He^4$ baths, so to keep completely disconnected the gravitationally interacting masses of superfluid. Otherwise, they would be rigidly connected in phase by the infinite range of superfluid order ODRLO [16].

We use only $\phi = \phi_1$ of eq(1), because d = d1 ≪ d2 = 2L. The mass m is given by $m = L\sigma\rho_s$. For the time $\Delta t$ it can be taken the characteristic time $\Delta t_J = 1/2f_J$, where $f_J$ is the Josephson frequency used to probe the phase [14]. In the ideal non-dissipative Josephson regime [17], which we consider here, this time is fundamental in that it marks the period with which the superfluid density (not the superfluid velocity as in the dissipative phase slippage regime) goes momentarily to zero. This happens when, during the Josephson current oscillations, the phase difference across the junction passes through $\Delta\phi = \pi$, see discussion in [16].

The relation to calculate the effects of QG in our version, as it comes from eq (1), is then

(2)    $\phi = A (L\sigma\rho_s)^2 (G/\hbar) (1/2f_J) (1/d)$

where A is a form factor of order O(1), which takes in account that in our case the masses are cylinders, while in [1,2] the masses are point-like or (nano)spheres. To evaluate the factor A, we take the eq (8) from a calculation [18] for the Coulomb potential per unit length between two charged isolated cylinders, and substitute mass to charge and the gravitational constant G in place of the electrostatic constants. With our parameters, A comes out to be around 0,5 without considering the finite size of the capillary radius dimensions.

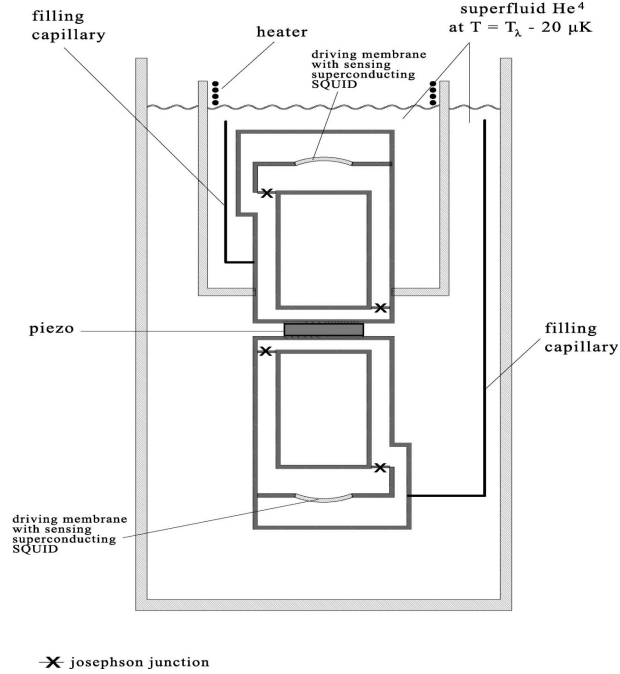

Fig.1: two SHeQUIDs, with the planes of the loops sensitive to the Earth rotation lay, in a vertical plane, with the horizontal channels of length L parallel, and are oriented to maximize the Earth rotation signal; for details of the superflow driving/measuring membrane assembly, including an electrostatic actuator and a superconducting SQUID, see ref[14]; for details of the Josephson junctions realized with submicron channels, see text and related refs; the SHeQUIDs are immersed in two separated liquid He$^4$ baths at T = $T_\lambda$ – 20 μK; each one is connected to its bath by a filling capillary; the side-by-side channels are made of dielectric material in view of the test as in Sec. 4; the piezo modulates the distance d between the side-by-side arms; in the QG case, each of the outputs of the two SHeQUIDs would be modulated at the piezo frequency; the heater burns the superfluid film to keep the two baths disconnected, and thus, when switched off, a putative QG effect should disappear, see text

The superfluid density $\rho_s$ has a definite temperature dependence $\rho_s(T) = 2.4\, \rho_\lambda\, (1 - T/T_\lambda)^{2/3}$ with $\rho_\lambda = 1.5\, 10^2$ Kg/m$^3$. As for the dimensions of the superfluid interferometer and for the realization of the Josephson junctions, the literature is abundant of elegant experiments [13,14]. So for a practical realization of our proposal, we suggest to use the typical realizations one can find therein: i) for the junctions, use arrays of hundreds of submicron channels in parallel in a few microns square lattice on a plate of submicron thickness, ii) for the channel cross section $\sigma$ and length L respectively $\sigma = 4\, 10^{-6}$ m$^2$ and L = $3\, 10^{-2}$ m, and, as the typical Josephson frequency used to probe the phase $f_J$ ranges between a fraction of 1 KHz to some 10 kHz [14], we take 5 kHz. For d we take d = $10^{-2}$ m. We fix for convenience the working temperature at about 20 μK below $T_\lambda$ - the λ

point - where the Josephson junctions are well in the ideal non-dissipative Josephson regime (in contrast to the dissipative phase slip regime farther below the λ point [17]). Typically, the temperature in these experiments is regulated with a stability of about 50 nK.

The experiment consists in testing if interference effects in SHeQUID are destroyed. The sign of *classical* response of the SHeQUID is the response to the Earth rotation. In particular by properly orienting around the vertical the SHeQUID depicted in Fig. 1, we can get the maximum interference signal [14]. Then it is straightforward to check if the gravitational interaction may destroy such an interference. In fact Eq (2) shows a strong dependence of the phase $\phi$, which regulates the interference effects on each SHeQUID, on the distance d. Modulating this distance, say by microns at Hz frequencies using a piezoelectric actuator and taking advantage of the elasticity of the apparatuses, $\phi$ can be modulated, for a $\delta d$ of 1 μm, to order of 1 rad. So the whole range of relevant phases can be explored, and, looking if in some conditions related to eq (2) the interference pattern is destroyed, it can be assessed uniquivocally if entanglement did indeed occurs.

4. A test of the proof of principle.
There are two requirements for a correct implementation of our proposal. First, the portions of superfluid involved - those traversing the SHeQUID interferometer - must be in N00N states [8], and, second, the Josephson junctions must indeed act as beam splitters, in order to consider the SHeQUID exactly similar to a Mach-Zehnder interferometer.

N00N states are expected to exist for the *mesoscopic* nanoparticles traversing the interferometers, as in [2]. The argument is that there is not enough energy to create excitations of the same mass, nor to dissociate the nanoparticles in components [8]. The situation for superfluid $He^4$ appears to be quite similar: as a consequence of ODLRO [16], all the atoms in a *macroscopic* sample are in a single macroscopic entangled state completely determined by the local fields [19].

It is not the target of this paper to dicuss theoretically this issue, nor the one concerning the operation of the Josephson junctions as beam splitters. Rather, we propose to clarify both issues via a preliminary test.

Let us consider an electromagnetic version [20] of the scheme above. Apply to the same apparatus, prepared for the proposed QG vs CG

experiment, a static electric field E, oriented horizontally in the plane of the figure . In the parallel side-by-side dielectric channels, the superfluid will polarize and a repulsive electrostatic force $F_E$ will arise. The dielectric constant of He4 is known [21], and classical electrostatics will give $F_E$, either in closed form, or numerically, if fringe effects are important. This will allow to estimate the field E to apply, for the experiment to be sensitive enough to give a yes/no answer. Now the classical em field of course emerges from a quantum field, and thus the effects of entanglement, now via the em field, must be found. This will uniquely happen, only if *both* the requirements above are fulfilled, and makes this preliminary test a suitable proof of principle for the gravitational case. As for the expected effects, we can get the relevant phases from a crude estimate, by substituting in eq (1) the electrostatic repulsion $F_E$ in place of the gravitational attraction $Gm^2/d^2$.

Of course, in performing such a preliminary experiment, if QG applies one has both contributions in the effects. Also, as the gravitational one arises from an attraction, while the electrostatic one comes from a repulsion, the two will have opposite signs, and there will be one $F_E$ which compensates the unknown gravitational one, so that in the end it would provide a measurement with a different procedure. This is an additional bonus coming from the considerations above

5. Discussion .
The expected effect is large, $\phi$'s of order of rads. This is because in our versions the gravitationally interacting masses involved are many orders larger than in the schemes of [1,2], $10^{-8}$ kg vs $10^{-12}$ - $10^{-14}$ kg respectively.

A necessary condition to observe the putative entanglement is to keep completely disconnected the interacting masses of superfluid, as noted in Sec 3. A connection may easily occur via the superfluid film. As it is well known, the film climbs any wall, which stays below $T_\lambda$, and spills over on the opposite side, effectively locking the baths to the same phase. To stop the film one burns it out with appropriate heaters above the surface, as in Fig1. This method will ensure that the baths are not connected.

Actually, this feature offers a yes/no test of uniquely compelling evidence. If the experiment gives a positive outcome, indicating the occurrence of entanglement, one can make sure of the conclusion by switching off the heater. If the effect disappears, while nothing else has been changed, this will be proof of the entanglement effect.

A few extraneous interaction must be examined, which may connect the masses. After usual shielding procedures with Faraday cages and mu-metal, em interferences/interactions should convincingly be excluded. Still one may be concerned with Casimir interactions. The distance d of 1 cm should be plenty to avoid the Casimir effects discussed in [22].

A disturbance on phases in the SHeQUIDS comes from the possible presence of quantized vortexes. Vortexes may be created in turbulent episodes during cooling below the temperature of transition to superfluidity $T_\lambda$. They may move and/or be metastable, giving occasional, abrupt and uncontrolled overall phase changes in the system. Fortunately such episodes occur at intervals of hours, see in particular Fig. 3 in [23], which shows long term drifts below $2 \times 10^{-3}$ rad over 6 hours. So, over the characteristic times of an actual experiment, this feature would not give any problem.

Another source of external disturbances in this type of experiments has been analyzed in [24] and concerns acceleration noise affecting the masses of the proposals [1,2]. In our version, this would not apply of course, but it would intervene another disturbance, now connected with uncontrolled rotational movements of the platform on which the whole experimental set up resides. The $He^4$ SQUID used here is sensitive to picking up the component the rotation of Earth over its sensitive area, and in fact the instrument is oriented, still in a vertical plane, to maximize such a pick up, in order to maximize its response. As discussed in [23], concerning the interest of $He^4$ SQUIDs as gyroscopes, this disturbance could have been greatly mitigated already at that time, see ref [16] therein. Since then there has been continuing progress in demonstrating rotationally ultra-quiet platforms, motivated by geophysical research [25] and towards laboratory tests of the Lense-Thirring effect [26]. Thus it should be feasible to go well beyond the requirements for the experiments proposed here.

To enter the theoretical debate on the significance of QG vs CG tests (3-5) is beyond the scope of this experimentally tuned work. However the following is of relevance here. A recent new perspective [10] has been worked out within the framework of Generalized Probabilistic Theories. At variance with previous debates, it has the advantage that it does not presuppose the quantum formalism. The result is that if gravity does not violate two conditions - i) faster-than-light signalling are not allowed and ii) of being a mediator which interacts only locally - then a positive outcome of an experiment like those of refs [1,2] would demonstrate *only* that gravity is *non-classical*. Therefore, as the present proposed experiment belongs to the same class of those of refs [1,2], the same conclusion applies.

5. Concluding remarks.

Other possible schemes would be based on altering the velocity of the superfluid, as refs in [14], where one would effectively alter the interaction time between the masses in the channels. Also, Josephson effects are similarly shown when the SHeQUID would work in the dissipative phase slippage regime. Either of the above may offer interesting possibilities, but extensions like these are beyond the scope of this note.

In conclusion, should the test of the proof of principle above give a positive answer, then it appears that the proposed experiment would give a yes or no test. If it would be strictly on QG vs CG, or limited to *non-classicity* vs *classicality*, it would depend on further developments of the debate. At the moment it is remarkable that the proposed experiment appears quite feasible, using only well demonstrated methods and technologies, with no need to extensions beyond current frontiers.

Acknowledgements

We enjoyed lively discussions in person with Alessandro Bettini. MC is indebted to Alessio Belenchia for the critical readings of the manuscript in the various versions and for constructive and encouraging comments. We thank Richard Packard and Andrea Vinante for useful criticisms and comments. Comments from Michele Bonaldi, Andrea Di Biagio, Giulio Gasbarri, Flaminia Giacomini and Carlo Rovelli were appreciated. We thank Antonello Ortolan for help in considering the form factor and Mario Zago for producing the figure. We thank an anonymous referee for suggestions to improve the writing.